\documentclass[aps,10pt,a4paper,twocolumn,showpacs,noshowkeys,notitlepage]{revtex4}

\usepackage{amsmath}
\usepackage{times}
\usepackage{amssymb}
\usepackage[dvipdfm]{graphicx}
\usepackage{hyperref}

\begin{document}

\title{Entanglement swapping of atomic states through the photonic Faraday
rotation}
\author{W.P. Bastos, W.B. Cardoso, A.T. Avelar, and B. Baseia}
\affiliation{Instituto de F\'{i}sica, Universidade Federal de Goi\'{a}s, 74.001-970, Goi\^{a}nia,
Goi\'{a}s, Brazil}

\begin{abstract}
We propose an entanglement swapping of atomic states confined by cavities QED using a photonic Faraday rotation. Two schemes are considered in which we use three and four cavities, respectively, plus an additional circularly-polarized photon. After interacting with an atom trapped inside the cavity the system evolves to an entangled atom-photon state. The entanglement swapping is then achieved by a Bell-state measurement upon the entire atom-photon state.
\end{abstract}


\maketitle

In quantum computation and quantum communication \cite{Nielsen00}
the entanglement performs a fundamental resource for many protocols,
such as quantum teleportation \cite{BennettPRL93}, quantum dense
coding \cite{Nielsen00}, and distributed quantum computation
\cite{BuzekPRA97}.  In this context, entanglement swapping plays an important role in several protocols of quantum information transfer, in particular, it is arguably one of the most important ingredients for quantum repeaters and quantum relays \cite{ZukowskiPRL93,BriegelPRL98}, as well as in teleportation of  entangled states. 
For the entanglement swapping protocol, two pairs of particles are usually used with each pair previously entangled, and a
Bell-state measurement made upon a particle of each pair leads the remaining particles to an entangled state, even if they have never
interacted previously with each other.

Due to the importance of entanglement swapping various experimental
results have been presented recently
\cite{PanPRL98,BoulantPRA03,JiaPRL04,Riedmatten,TakeiPRL05,GoebelPRL08,RiebeNP08,LuPRL09}.
In Ref. \cite{PanPRL98}  the authors employed two pairs of polarized entangled photons and making a Bell-state measurement upon the photon of each pair, they displayed an entanglement of freely propagating particles that have never interacted or dynamically coupled by any other means. Entanglement swapping operations have been reported in \cite{BoulantPRA03} via nuclear magnetic resonance quantum-information processing, over long distances in optical fibers \cite{Riedmatten}, and  unconditionally for continuous variables  \cite{JiaPRL04}.  The entanglement swapping for continuous-variable has been used to realize quantum teleportation beyond the no-cloning limit \cite{TakeiPRL05}. Multistage entanglement swapping in photonic system \cite{GoebelPRL08}, an ion-trap quantum processor through entanglement swapping \cite{RiebeNP08}, and  the first experimental demonstration of the Greenberger-Horne-Zeilinger entanglement swapping \cite{LuPRL09} have also been reported. 

On the other hand, a lot of theoretical schemes have appeared in the
recent literature \cite{BosePRA98,BoudaJPA01,KarimipourPRA02,LiPLA04,LeePRA04,ZhangPLA05,ZhangPRA05,ZhanOC09} generalizing the standard entanglement swapping for: multiparticle systems \cite{BosePRA98}; multi-qudit  systems \cite{BoudaJPA01}, where the authors have extended the scheme originally proposed for two pairs of qubits and an arbitrary
number of systems composed by an arbitrary number of qudits;  d-level systems in a generalized cat
state \cite{KarimipourPRA02}, useful for protocol of secret sharing.  Concerning with secret sharing in cavity QED \cite{ZhangPLA05}, multiparty secret sharing of quantum information \cite{LiPLA04} and classical information \cite{ZhangPRA05}, secure multiparty quantum communication by Bell states \cite{LeePRA04} and by entangled qutrits  \cite{ZhanOC09} based on entanglement swapping have also been proposed. 

In addition to the list of entanglement swapping applications, the
purification of entangled states by local actions, using a variant
of entanglement swapping, was studied in Refs.
\cite{BosePRA99,ShiPRA00,YangPRA05-2}. This issue was extended for  continuous variables in \cite{PolkinghornePRL99}. The quantum key distribution schemes  \cite{SongPRA04} and  teleportation of a two-particle entangled state \cite{LuPLA00,CardosoPLA09} employing entanglement swapping have been reported, as well as  entanglement swapping without joint measurement \cite{YangPRA05}. In the QED-cavity
context, a scheme based on two atoms and two cavities initially
prepared in two pairs of atom-photon nonmaximally entangled states,
was considered in \cite{WuCTP06} to create maximally entangled
photon-photon and atom-photon states via entanglement swapping, with
atomic states in either a three-level cascade or lambda
configuration in \cite{GuerraJMO06}, with resonant interaction of a
two-mode cavity with a $\lambda$-type three-level atom involving
only a single measurement in \cite{YangCTP07} and, in
\cite{HeIJQI07}, an alternative scheme to implement the entanglement
swapping. More recently, an entanglement swapping in the
two-photon Jaynes-Cummings model was proposed in \cite{dSouzaPS09}. 

\begin{figure}[t]
\includegraphics[width=2.8cm]{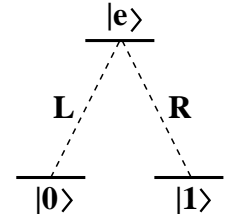}
\caption{Atomic configuration of the three-level atom trapped in the low-Q cavities.}
\label{F1}
\end{figure}

Here, taking advantage of the quantum regime of strong interactions between single atoms and photons present in a microtoroidal resonator \cite{DayanSCI08}, as employed in \cite{AnPRA09} for quantum information processing, we propose an entanglement swapping of states of atoms confined in distant low QED cavities using photonic Faraday rotations. The main idea is to make use of the Faraday rotation produced by single-photon-pulse input and output process regarding low-Q cavities \cite{JulsgaardNAT01}.  
In view of our applications, we revisited the input-output relation for a cavity coherently interacting with a trapped two-level atom, recently considered in Ref. \cite{AnPRA09}. We consider a three-level atom interacting with a single mode of a low-Q cavity pumped by photonic emission of a single photon source via optical fibers. Fig. \ref{F1} shows the atomic levels of the atom trapped inside the cavities. Each transition is described by the Jaynes-Cummings model.

One can use the quantum Langevin equation of the cavity
mode $a$ driven by the corresponding cavity input operator $a_{\mathrm{in}%
}(t)$ and the atomic lowering
operator are \cite{Walls94},
\begin{subequations}
\begin{eqnarray}
\dot{a}(t)=-[i(\omega _{c}-\omega _{\mathrm{p}})+\frac{\kappa }{2}%
]a(t)-g\sigma _{-}(t)-\sqrt{\kappa }a_{\mathrm{in}}(t),
\label{eom}
\end{eqnarray}%
\begin{eqnarray}
\dot{\sigma}_{-}(t)&=&-[i(\omega _{0}-\omega _{\mathrm{p}})+\frac{\gamma }{2}%
]\sigma _{-}(t) \nonumber \\ &-&g\sigma _{z}(t)a(t)+\sqrt{\gamma }\sigma _{z}(t)b_{\mathrm{in%
}}(t),
\label{eom2}
\end{eqnarray}%
\end{subequations}
respectively, where $a$ $(a^{\dag })$ is the annihilation (creation) operator of the
cavity field with frequency $\omega _{c}$; $\sigma _{z}$ and $\sigma _{+}$ ($\allowbreak \sigma _{-}$) are, respectively, inversion and raising (lowering) operators of the two-level atom with frequency
difference $\omega _{0}$ between these two-levels. $\kappa $ and $\gamma $ are, respectively, the cavity damping rate and the atomic decay rate. The vacuum input field $b_{\mathrm{in}}(t)$  felt by the two-level atom satisfies the commutation relation $[b_{\mathrm{in}}(t),b_{\mathrm{in}}^{\dag }(t^{\prime })]=\delta (t-t^{\prime })$. The input and output fields of
the cavity are related by the intracavity field as $ a_{\mathrm{out}}(t)=a_{\mathrm{in}}(t)+\sqrt{\kappa }a(t)$ \cite {Walls94}.

In this way, considering a large enough $\kappa$ to be sure  that we have a weak excitation by the single-photon pulse on the atom initially prepared in the ground state, i.e., keeping $\langle \sigma_{z}\rangle =-1$ throughout our operation, as shown in \cite{AnPRA09} one can adiabatically eliminate the cavity mode and arrive at the input-output relation of the cavity field,%
\begin{equation}
r(\omega _{\mathrm{p}})=\frac{[i(\omega _{c}-\omega _{\mathrm{p}})-\frac{%
\kappa }{2}][i(\omega _{0}-\omega _{\mathrm{p}})+\frac{\gamma }{2}]+g^{2}}{%
[i(\omega _{c}-\omega _{\mathrm{p}})+\frac{\kappa }{2}][i(\omega _{0}-\omega
_{\mathrm{p}})+\frac{\gamma }{2}]+g^{2}},  \label{rela}
\end{equation}%
where $r(\omega _{\mathrm{p}})\equiv a_{\mathrm{out}}(t)/a_{\mathrm{in%
}}(t)$ is the reflection coefficient of the atom-cavity system. Now, considering the case
of $g=0$ and an empty cavity we have \cite{Walls94}
\begin{equation}
r_{0}(\omega _{\mathrm{p}})=\frac{i(\omega _{c}-\omega _{\mathrm{p}})-\frac{%
\kappa }{2}}{i(\omega _{c}-\omega _{\mathrm{p}})+\frac{\kappa }{2}}.
\label{r0}
\end{equation}

According to \cite{AnPRA09} the transitions $|e\rangle
\leftrightarrow |0\rangle $ and $|e\rangle \leftrightarrow |1\rangle
$ are due to the coupling to two degenerate cavity modes
$a_{\mathrm{L}}$ and $a_{\mathrm{R}}$ with left (\textrm{L}) and
right (\textrm{R}) circular polarization, respectively. For the atom initially prepared in $|0\rangle $, the only possible transition $|0\rangle \rightarrow |e\rangle $ implies that only the \textrm{L} circularly polarized single-photon pulse
$|\mathrm{L}\rangle $ will work. Hence Eq. (\ref{rela}) leads the input pulse to the output one as $%
|\Psi _{\mathrm{out}}\rangle _{\mathrm{L}}=r(\omega _{\mathrm{p}})|\mathrm{L}%
\rangle \approx e^{i\phi }|\mathrm{L}\rangle $ with $\phi $ the
corresponding phase shift being determined by the parameter values.  
Note that an input \textrm{R} circularly polarized
single-photon pulse $|\text{R}\rangle $ would only sense the empty
cavity; as a consequence the corresponding output governed by Eq. (\ref{r0}) is $|\Psi _{\mathrm{out}}\rangle _{\mathrm{R}%
}=r_{0}(\omega _{\mathrm{p}})|\mathrm{R}\rangle =e^{i\phi _{0}}|\mathrm{R}%
\rangle $ with $\phi_{0}$ a phase shift different from $\phi$. Therefore,
for an input linearly polarized photon pulse $|\Psi _{\mathrm{in}}\rangle =%
\frac{1}{\sqrt{2}}(|\mathrm{L}\rangle +|\mathrm{R}\rangle )$, the
output pulse is
\begin{equation}
|\Psi _{\mathrm{out}}\rangle _{-}=\frac{1}{\sqrt{2}}(e^{i\phi }|\mathrm{L}%
\rangle +e^{i\phi _{0}}|\mathrm{R}\rangle ).
\end{equation}%
This also implies that the polarization direction of the reflected photon rotates an angle $\Theta _{F}^{-}=(\phi_0 -\phi)/2$ with respect to that of the input one, called Faraday rotation \cite{JulsgaardNAT01}. If the atom is initially prepared in $|1\rangle $, then only the $\mathrm{R}$ circularly polarized photon could sense the atom, whereas the $\mathrm{L}$ circularly polarized photon only interacts with the empty cavity. So we have,
\begin{equation}
|\Psi _{\mathrm{out}}\rangle _{+}=\frac{1}{\sqrt{2}}(e^{i\phi _{0}}|\mathrm{L}\rangle +e^{i\phi }|\mathrm{R}\rangle ),
\end{equation}
where the Faraday rotation is $\Theta _{F}^{+}=(\phi -\phi _{0})/2$.


\begin{figure}[t]
\includegraphics[width=5cm]{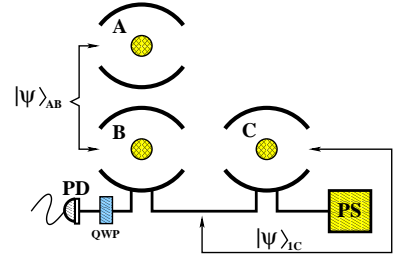}
\caption{(Color online) Schematic diagram of the entanglement swapping procedure for the case 1 using three three-level atoms trapped in cavities $A$, $B$, and $C$, a single photon source (PS), a quarter-wave plate (QWP), and a photodetector of polarization (PD).}
\label{F2}
\end{figure}

\textit{Case $\#$ 1} - Firstly, we assume the previously entangled state between the atoms confined in the cavities $A$ and $B$, given by
\begin{equation}
 |\psi\rangle_{AB}=\frac{1}{\sqrt{2}}(|01\rangle_{AB}+|10\rangle_{AB}).
\end{equation}
In another spatial position, an entanglement has been previously prepared in atom confined in cavity $C$ with a photon, named $1$, in Faraday rotated state, written in the form
\begin{equation}
 |\psi_{C1}\rangle= \frac{1}{\sqrt{2}}(|0\rangle_C|\eta_{1}\rangle_{-}+|1\rangle_C|\eta_{1}\rangle_{+}),
\end{equation}
where $ |\eta_{1}\rangle_{-}=(e^{i\phi}|L\rangle_1+e^{i\phi_{0}}|R\rangle_1)/\sqrt{2}$ and  $|\eta_{1}\rangle_{+}=(e^{i\phi_{0}}|L\rangle_1+e^{i\phi}|R\rangle_1)/\sqrt{2}$.
This entanglement is constructed via interaction of the polarized photon (in the state $|\psi\rangle_1=(|L\rangle_1+|R\rangle_1)/\sqrt{2}$) with a three-level atom (previously prepared in $|\psi\rangle_C=(|0\rangle_C+|1\rangle_C)/\sqrt{2}$). Fig. \ref{F2} shows the entire procedure of the case 1.

The entanglement swapping is realized through a Bell-state measurement on the system composed by the photon and the atom confined in the cavity $B$. To this end, we send the photon through the cavity $B$ to interact with the atom, leading the state of the whole system given as
\begin{eqnarray}
|\psi^{\prime}\rangle&=&\frac{1}{2\sqrt{2}}\left[|00\rangle_{AC}(e^{i(\phi+\phi_{0})}|L\rangle_1+e^{i(\phi+\phi_{0})}|R\rangle_1)|1\rangle_{B} \right. \nonumber\\
&+&|01\rangle_{AC}(e^{2i\phi_{0}}|L\rangle_1+e^{2i\phi}|R\rangle_1)|1\rangle_{B}\nonumber\\
&+&|10\rangle_{AC}(e^{2i\phi}|L\rangle_1+e^{2i\phi_{0}}|R\rangle_1)|0\rangle_{B}\nonumber\\
&+&\left. |11\rangle_{AC}(e^{i(\phi+\phi_{0})}|L\rangle_1+e^{i(\phi+\phi_{0})}|R\rangle_1)|0\rangle_{B} \right].
\end{eqnarray}
Next, assuming $\phi=\pi$ and $\phi_{0}=\pi/2$, plus the application of a Hadamard operation upon the state of the atom $B$ and the photon via an external laser beam and a quarter-wave plate (QWP), respectively, the state of the entire system evolves to
\begin{eqnarray}
 |\psi^{\prime\prime}\rangle&=&\frac{1}{2\sqrt{2}}\left[-i|L0\rangle_{1B}(|00\rangle_{AC}+|11\rangle_{AC}) \right. \nonumber\\
&+&i|L1\rangle_{1B}(|00\rangle_{AC}-|11\rangle_{AC})\nonumber\\
&-&|R0\rangle_{1B}(|01\rangle_{AC}-|10\rangle_{AC})\nonumber\\
&+&\left. |R1\rangle_{1B}(|01\rangle_{AC}+|10\rangle_{AC}) \right].
\end{eqnarray}
Finally, appropriate detections of the state of the atom $B$ and the polarized photon state conclude the entanglement swapping. Table \ref{T1} summarizes the atomic rotations to complete the entanglement swapping.

\begin{table}
\begin{center}
\begin{tabular}{|| c | c | c ||}
\hline  MAPS                  &  ESR                                                          & AO \\
\hline  $|L0\rangle_{1B}$ &  $|00\rangle_{AC}+|11\rangle_{AC}$ & $\sigma_x$ \\
\hline  $|L1\rangle_{1B}$ &  $|00\rangle_{AC}-|11\rangle_{AC}$  & $i\sigma_y$ \\
\hline  $|R0\rangle_{1B}$ &  $|01\rangle_{AC}-|10\rangle_{AC}$  & $\sigma_z$ \\
\hline  $|R1\rangle_{1B}$ &  $|01\rangle_{AC}+|10\rangle_{AC}$ & $\mathbb{I}$ \\
\hline
\end{tabular}
\end{center}
\caption{Atomic rotations completing the entanglement swapping procedure for the case 1. The first column represents the measurement in the atom $B$ and the photon states (MAPS), second column is the result of entanglement swapping resulting (ESR), and the third is the atomic operation (AO) considering the atom $A$ (local) represented by Pauli operators with $\mathbb{I}$ being the identity operator.}
\label{T1}
\end{table}

\begin{figure}[t!]
\includegraphics[width=5cm]{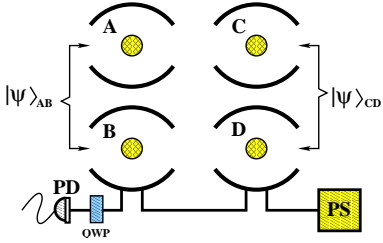}
\caption{(Color online) Schematic diagram of the entanglement swapping procedure for de case 2, using the same notation of Fig. \ref{F2} plus an additional atom confined in cavity $D$.}
\label{F3}
\end{figure}

\textit{Case $\#$ 2} - Now, we start by considering two pairs of atoms, trapped inside the cavities $A$, $B$, $C$, and $D$, and previously entangled in the following states
\begin{eqnarray}
&&|\psi\rangle_{AB}=\frac{1}{\sqrt{2}}(|01\rangle_{AB}+|10\rangle_{AB}), \nonumber \\
&&|\psi\rangle_{CD}=\frac{1}{\sqrt{2}}(|01\rangle_{CD}+|10\rangle_{CD}).
\end{eqnarray}
The scheme for entanglement swapping is summarized in Fig. \ref{F3}.

Firstly, an auxiliary photon is sent to interact with the atom confined in the cavity $D$, leading the state of the entire atom-photon system to the form
\begin{eqnarray}
 |\varphi\rangle&=&\frac{1}{2}(|01\rangle_{AB}+|10\rangle_{AB}) \nonumber \\
 &\otimes&(|01\rangle_{CD}|\eta_{1}\rangle_{+}+|10\rangle_{CD}|\eta_{1}\rangle_{-}).
\end{eqnarray}
Next, considering a Hadamard operation upon the atom $D$, we have
\begin{eqnarray}
|\varphi^{\prime}\rangle&= &\frac{1}{4}\left[e^{i\phi_{0}}|L010\rangle_{ABC}(|0\rangle_{D}-|1\rangle_{D}) \right. \nonumber\\
&+& e^{i\phi}|L011\rangle_{ABC}(|0\rangle_{D}+|1\rangle_{D})\nonumber\\
&+&e^{i\phi}|R010\rangle_{ABC}(|0\rangle_{D}-|1\rangle_{D})\nonumber\\
&+&e^{i\phi_{0}}|R011\rangle_{ABC}(|0\rangle_{D}+|1\rangle_{D})\nonumber\\
&+&e^{i\phi_{0}}|L100\rangle_{ABC}(|0\rangle_{D}-|1\rangle_{D})\nonumber\\
&+&e^{i\phi}|L101\rangle_{ABC}(|0\rangle_{D}+|1\rangle_{D})\nonumber\\
&+&e^{i\phi}|R100\rangle_{ABC}(|0\rangle_{D}-|1\rangle_{D})\nonumber\\
&+& \left. e^{i\phi_{0}}|R101\rangle_{ABC}(|0\rangle_{D}+|1\rangle_{D})\right].
\end{eqnarray}
In sequence, the photon emerging form the cavity $D$ is sent to interact with the atom in the cavity $B$ and, soon after, a Hadamard operation upon the atom $B$ and upon the photon states, transforms the state of the whole system to the form

\begin{widetext}
\begin{eqnarray}
 |\varphi^{\prime\prime}\rangle&=&\frac{1}{8} \left\{ [(e^{2i\phi}+e^{2i\phi_{0}})|L\rangle_1-(e^{2i\phi}-e^{2i\phi_{0}})|R\rangle_1]|00\rangle_{AC}(|00\rangle_{BD}-|01\rangle_{BD}-|10\rangle_{BD}+|11\rangle_{BD}) \right. \nonumber\\
&+&[(e^{2i\phi}+e^{2i\phi_{0}})|L\rangle_1+(e^{2i\phi}-e^{2i\phi_{0}})|R\rangle_1]|11\rangle_{AC}(|00\rangle_{BD}+|01\rangle_{BD}+|10\rangle_{BD}+|11\rangle_{BD})\nonumber\\
&+&2e^{i(\phi+\phi_{0})}|L\rangle_1|01\rangle_{AC}(|00\rangle_{BD}+|01\rangle_{BD}-|10\rangle_{BD}-|11\rangle_{BD})\nonumber\\
&+& \left. 2e^{i(\phi+\phi_{0})}|L\rangle_1|10\rangle_{AC}(|00\rangle_{BD}-|01\rangle_{BD}+|10\rangle_{BD}-|11\rangle_{BD}) \right\},
\end{eqnarray}
and with the single choice $\phi=\pi$ and $\phi_{0}=\pi/2$, we obtain
\begin{eqnarray}
 |\varphi^{\prime\prime\prime}\rangle&=&\frac{1}{4}\left[|R00\rangle_{BD}(|11\rangle_{AC}-|00\rangle_{AC})+|R01\rangle_{BD}(|11\rangle_{AC}+|00\rangle_{AC}) \right. \nonumber\\
&+&|R10\rangle_{BD}(|11\rangle_{AC}+|00\rangle_{AC})+|R11\rangle_{BD}(|11\rangle_{AC}-|00\rangle_{AC})\nonumber\\
&-&i|L00\rangle_{BD}(|01\rangle_{AC}+|10\rangle_{AC})-i|L01\rangle_{BD}(|01\rangle_{AC}-|10\rangle_{AC})\nonumber\\
&+&\left. i|L10\rangle_{BD}(|01\rangle_{AC}-|10\rangle_{AC})+i|L11\rangle_{BD}(|01\rangle_{AC}+|10\rangle_{AC}) \right].
\end{eqnarray}
\end{widetext}
So, with a detection of the photon polarization plus separated atomic measurements in the atoms $B$ and $D$, one concludes the entanglement swapping. Table \ref{T2} presents the atomic operation to reconstruct the initial state.

\begin{table}
\begin{center}
\begin{tabular}{|| c | c | c ||}
\hline MAPS & ESR & AO \\
\hline  $|R00\rangle_{BD}$  & $|11\rangle_{AC}-|00\rangle_{AC}$ & $-i\sigma_y$ \\
\hline  $|R01\rangle_{BD}$  & $|11\rangle_{AC}+|00\rangle_{AC}$ & $\sigma_x$ \\
\hline $|R10\rangle_{BD} $ &  $|11\rangle_{AC}+|00\rangle_{AC}$ &  $\sigma_x$\\
\hline  $|R11\rangle_{BD}$  & $|11\rangle_{AC}-|00\rangle_{AC}$ &  $-i\sigma_y$\\
\hline $|L00\rangle_{BD} $ &  $|01\rangle_{AC}+|10\rangle_{AC}$ & $\mathbb{I}$ \\
\hline  $|L01\rangle_{BD}$  & $|01\rangle_{AC}-|10\rangle_{AC}$ & $\sigma_z$ \\
\hline  $|L10\rangle_{BD}$  & $|01\rangle_{AC}-|10\rangle_{AC}$ & $\sigma_z$ \\
\hline  $|L11\rangle_{BD}$  & $|01\rangle_{AC}+|10\rangle_{AC}$ & $\mathbb{I}$ \\
\hline
\end{tabular}
\end{center}
\caption{Atomic rotations completing the entanglement swapping procedure for the case 2. The first column represents the measurement in the atoms $BD$ and the photon states (MAPS); second column is the result of entanglement swapping resulting (ESR), and the third is the atomic operation (AO) considering the atom $A$ (local) represented by Pauli operators with $\mathbb{I}$ being the identity operator.}
\label{T2}
\end{table}

In conclusion, we presented an entanglement swapping of atomic
states confined in cavities QED using photonic Faraday rotation.
This scheme involves only virtual excitations of the atoms and the
cavities considered here are low-Q cavities, i.e., with large decay.
We also have considered the scheme with ideal detectors and fibers
without absorption. Two protocols for entanglement swapping were
demonstrated and in view of the current status of the technology our
scheme is feasible. This scheme can also be modeled in a quantum-dot
system, by simply replacing the atoms by excitons
\cite{LeuenbergerPRL05}.

\section*{Acknowledgement}

We thank the CAPES, CNPq, and FUNAPE-GO, Brazilian agencies, for the partial support.


\end{document}